# Gateway Placement and Selection Solutions in Wireless Mesh Networks: A survey

*Mohsen Jahanshahi[1], Arash Bozorgchenani[2,3]*

*[1]Dept. of Computer Engineering, Central Tehran Branch, Islamic Azad University, Tehran, Iran*
*[2] Dept. of Information Technology Engineering, Qazvin Branch, Islamic Azad University, Qazvin, Iran*
*[3] Dept. of Electrical, Electronic and Information Engineering, University of Bologna, Italy*

*mjahanshahi@iauctb.ac.ir,arash.bozorgchenani2@unibo.it*

**Abstract:** Due to the high demand of Internet access by users, and the tremendous success of wireless technologies, Wireless Mesh Networks (WMNs) have become a promising solution. IGW Placement and Selection (GPS) are significantly investigated problems to achieve QoS requirements, network performance, and reduce deployment cost in WMNs. Best effort is made to classify different works in the literature based on network characteristics. At first, one of the most principal capabilities of WMNs, which is taking advantage of using multi-radio routers in a multi-channel network, is studied. In this article, GPS protocols considering a definition of three types of WMN are investigated based on channel-radio association including Single Radio Single Channel (SRSC), Single Radio Multi- Channel (SRMC), and Multi-Radio Multi-Channel (MRMC) WMNs. Furthermore, a classification regarding static and dynamic channel allocation policies is derived. In addition, the reported works from the viewpoint of network solutions are classified. The first perspective is: centralized, distributed or hybrid architectures. Following this classification, the studies are categorized regarding optimization techniques, which are operation research-based solutions, heuristic algorithms, and meta-heuristic-based algorithms.

**Keywords**: Wireless mesh networks, IGW placement, IGW selection, multi-channel multi radio, optimization techniques

## 1. **Introduction**

Wireless Mesh Networks (WMNs) are a promising technology which have emerged since early 2000s and have received lots of attention. WMNs have certain merits which make them an economical solution for wireless broadband access. Self-healing, cheap-to-deploy and high scalability are characteristics of WMNs which have made this means of connectivity attractive for city projects and lots of application scenarios[1].

WMN nodes are comprised of two types: Mesh Routers (MRs) and Mesh Clients (MCs) that form a multi-hop wireless network connected to the Internet to provide end users with backhaul access [2]. MRs (a.k.a. mesh access routers), which have minimal mobility, form a wireless backbone that as well as providing wireless connections for MCs in their respective domains and acting as classical access points, relay each other’s packets in a multi-hop fashion. MCs can be fixed or mobile and they can be associated with one of the MRs to access Internet through Internet GateWays (IGWs) by multi-hop forwarding. IGWs are some special MRs which are configured with wired links and are directly connected to the Internet. Furthermore, they act as bridges between the WMN and the Internet. Figure 1 illustrates the structure of WMN components.

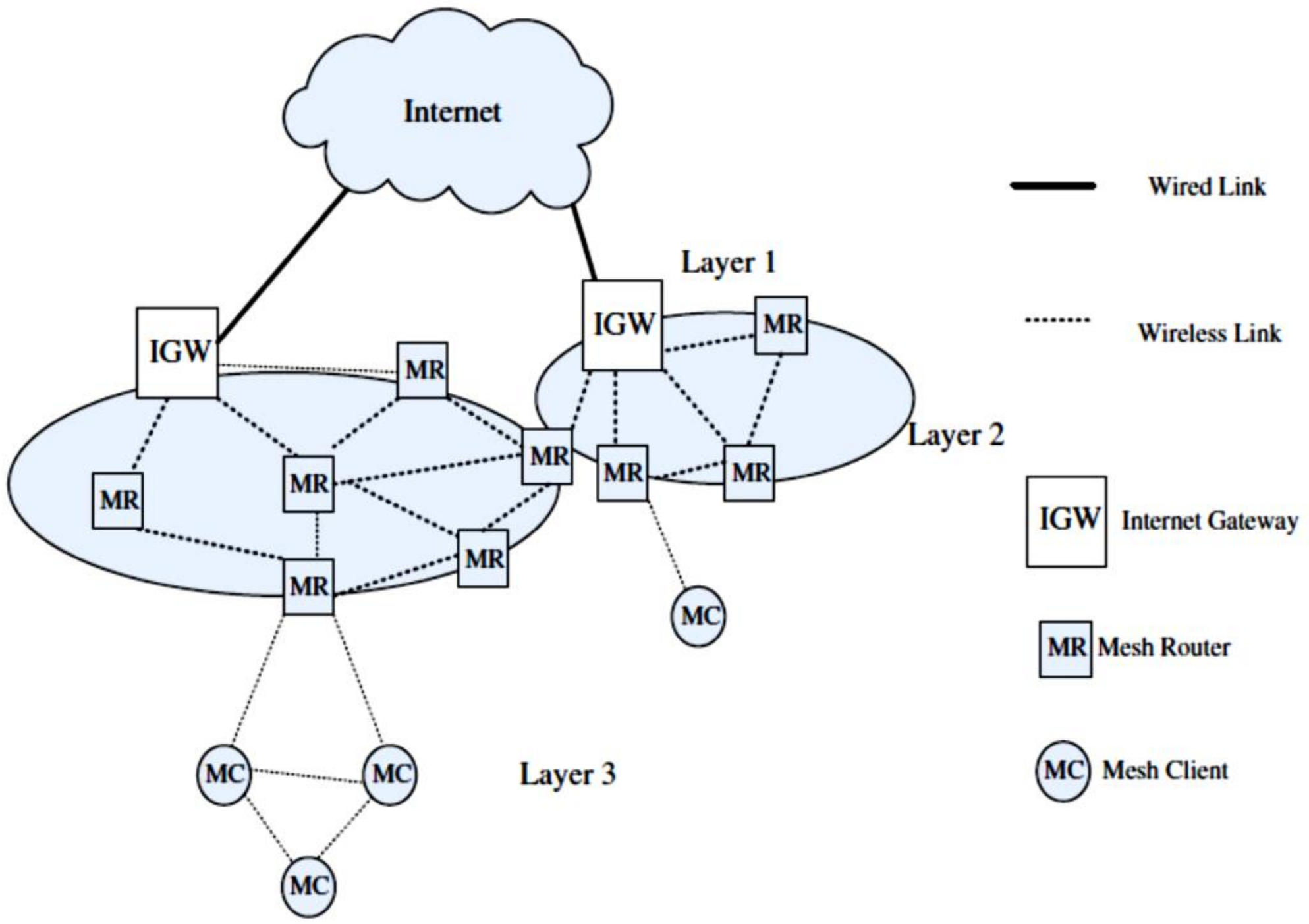

**Figure 1:** The architecture of a typical WMN[3]

Long distances are broken into a series of shorter hops by intermediate nodes, which not only sustain signal strength, but also forward packages on behalf of other nodes based on their knowledge of network. To deliver heterogeneous traffic, the optimization of the overall network performance of WMNs across multiple protocol layers is a critical issue which can be designed in either layered-protocol methodology or cross-layer methodology [2].

The architecture of WMNs is classified into 3 types: Infrastructure WMNs, Client WMNs, and Hybrid WMNs [1]. In an infrastructure or backbone WMN, the MRs form the backbone for the MCs and some of them have the capability to act as IGWs. Such an infrastructure provides backbone for MCs and enables integration of WMNs such as Wi-Fi networks, cellular networks, sensor networks with the Internet. The MCs themselves act as MRs in a Client WMN. By performing the tasks of routing, configuration functionality and providing end-user applications to customers they form the infrastructure of the network. Hybrid WMN is a combination of the two aforementioned architectures which means the MCs can access the network either through the MRs forming the backbone or through other MCs which forward the traffic to MRs. Besides, in this type, connectivity to other networks such as the Wi-Fi, WIMAX, cellular, and sensor networks is provided.

Placing the IGWs in a proper way is very crucial due to the additional expenses which increase the overall cost. IGW placement is the study of where to deploy the IGWs in order to reduce the network expenses and how to place them to satisfy QoS constraints. Loads of research has been carried out on IGW placement. However, some significant parameters have not yet been considered for this problem and need to be further investigated. Some of the major concerns are number of IGWs, which implies deployment cost, the average MR and IGW hop count which indicates the network delay, load balancing, interference, and loss ratio. Nevertheless, to the best of our knowledge parameters like reliability, which can have a great impact on the performance of the network, have not been taken into account in any of the works.

Apart from the subject of IGW placement, to have access to Internet, users need to be linked to a IGW whether directly or indirectly through an MR. For obtaining good communication quality, having a balanced network and covering all the nodes in the network, it is of vital importance to select an appropriate node to be given the IGW functionality. Therefore, loads of works have been dedicated to the problem of IGW selection. This problem, which is an NP problem [4], has a profound impact on the performance of WMN. To recover communication in the disaster area where some nodes are heavy-loaded and some are light-loaded, deployment of IGW in the center of the network cannot be done easily. As a consequence, there is a need to select an optimal node as the IGW to guarantee the maximum throughput in the network [5]. Some works have proposed a solution for selecting a MRs for relaying the packets from fixed and mobile users to mesh networks and on the other hand some have concentrated only on IGW nodes for relaying the packets from mesh network to Internet. In this survey, we have focused on both MRs selection and IGW selection.

IGW selection has been defined differently in various articles. In some articles it is defined as the process of selecting an appropriate node to be given the IGW functionality. This node can be selected based on some constraints and factors. While, in some other works the authors made an effort to select an IGW for routing the packets from the source node to the selected IGW. In other words, when packets are going to be sent from a node to next node, there should be some factors for the selection of the next node for relaying the packets. This type of IGW selection is out of our research scope since it is more concerned with routing and IGW discovery.

Lots of recent studies in WMNs have focused on multicasting [3, 6-10], link scheduling [11-14], channel assignment [7, 15-17], IGW placement, and IGW selection. Up to now, a lot of research in GPS has been conducted. Some of them concentrated on load balancing while some other works tried to take the interference into consideration by employing MRMC WMNs. On the other hand, some works proposed operation research-based methods and some studies argued heuristic and meta-heuristic-based solutions. However, few works considered some of these issues together in their studies. In other words, some works concentrated on clustering and some on IGW selection, but few have focused on both issues together. Since there is a tight relation between the topics of path selection, routing, clustering and channel assignment, considering these problems together and proposing a multi-objective optimization can greatly enhance network throughput. These are some of the problems which need to be investigated further and to this aim we are motivated to review different perspectives in GPS. In this article, related studies in GPS are investigated and classified into two main categories of network characteristics and network solutions. The former classification is done with regard to channel radio association as well as static or dynamic methods. The later classification is dedicated to optimization techniques, and centralized, distributed, or hybrid architectures. Our aim is to provide a better understanding of research challenges of this emerging and developed technology. Moreover, we hope that by analyzing different solutions in the literature and going through different approaches and their advantages and disadvantages, more investigation is done in the future for providing better services for the users and satisfying QoS requirements.

The rest of the article is structured as follows: In section 2, the preliminaries of GPS in WMNs are presented. Section 3 is dedicated to investigation of different classifications in GPS. An overall preview of some of the mostly related studies is summarized in section 4. This article concludes in section 5.

## 2. Preliminaries of GPS

In WMNs, placing the IGWs in a proper way is a key factor in terms of the optimal throughput, load balancing on the IGWs and satisfying QoS requirements. If IGWs are placed in areas with low traffic or few numbers of MCs they might be underutilized. That is why lots of works have been dedicated to the IGW placement problem in order to enhance the network performance. These works try to consider different parameters in placing IGWs in different areas. The more IGWs placed, the better performance will be gained, but, the higher the cost [18]. Therefore, there is a tradeoff between the number of IGWs which are placed and the QoS parameters.

The IGW placement problem is usually considered NP hard, thus near optimal heuristics are generally employed [4, 18-27]. The problem of formulations for promoting IGW placement is usually placed in one of the following categories:

- Single-objective: Optimization of IGW placement in some works has been done considering a single objective. Articles [24, 26, 28-32] can be categorized in this scope.

- Multi-objective: IGW placement can be optimized with respect to multiple objectives comprising throughput, delay, cost and so on. Studies such as [18, 19, 22, 25, 27, 33]optimized their works in a multi-objective fashion.

To optimize the IGW placement, the objective functions are usually bounded by several constraints for restricting the solutions to acceptable practical limits. A basic constraint for network flow problem formulation is the flow conservation constraint for balancing the total amount of in-flow and outward flow for the ideal link model [26, 29, 30].The objective function in some works is bounded by hop count constraint [22, 25, 31, 32]. The other constraint is IGW capacity (or cluster size) constraints that is measured by the maximum number of MRs that an IGW can serve [18, 19, 22, 27, 30-32]. Delay (or cluster radius) constraint is a main QoS constraint which is accumulated delay of communication hops between the MRs and their IGWs and should not be greater than a threshold [19, 32]. Relay load (or link load) constraints is also one of the principal QoS constraints which is an upper bound on the maximum number of MRs that can be transmitted through an individual MR in the neighborhood cluster [18, 19, 28-32]. Interference has also been considered a constraint in different studies [26, 27, 30].

## 3. Different Classifications in GPS Protocols

In this section, the mostly related studies in GPS are investigated. These works are classified in four perspectives: channel-radio association, static and dynamic channel assignment policies, centralized, distributed, and hybrid architectures and optimization techniques. The first two classifications are related to network characteristics and the second ones are related to network solutions. Different classifications are concisely depicted in Figure 2.

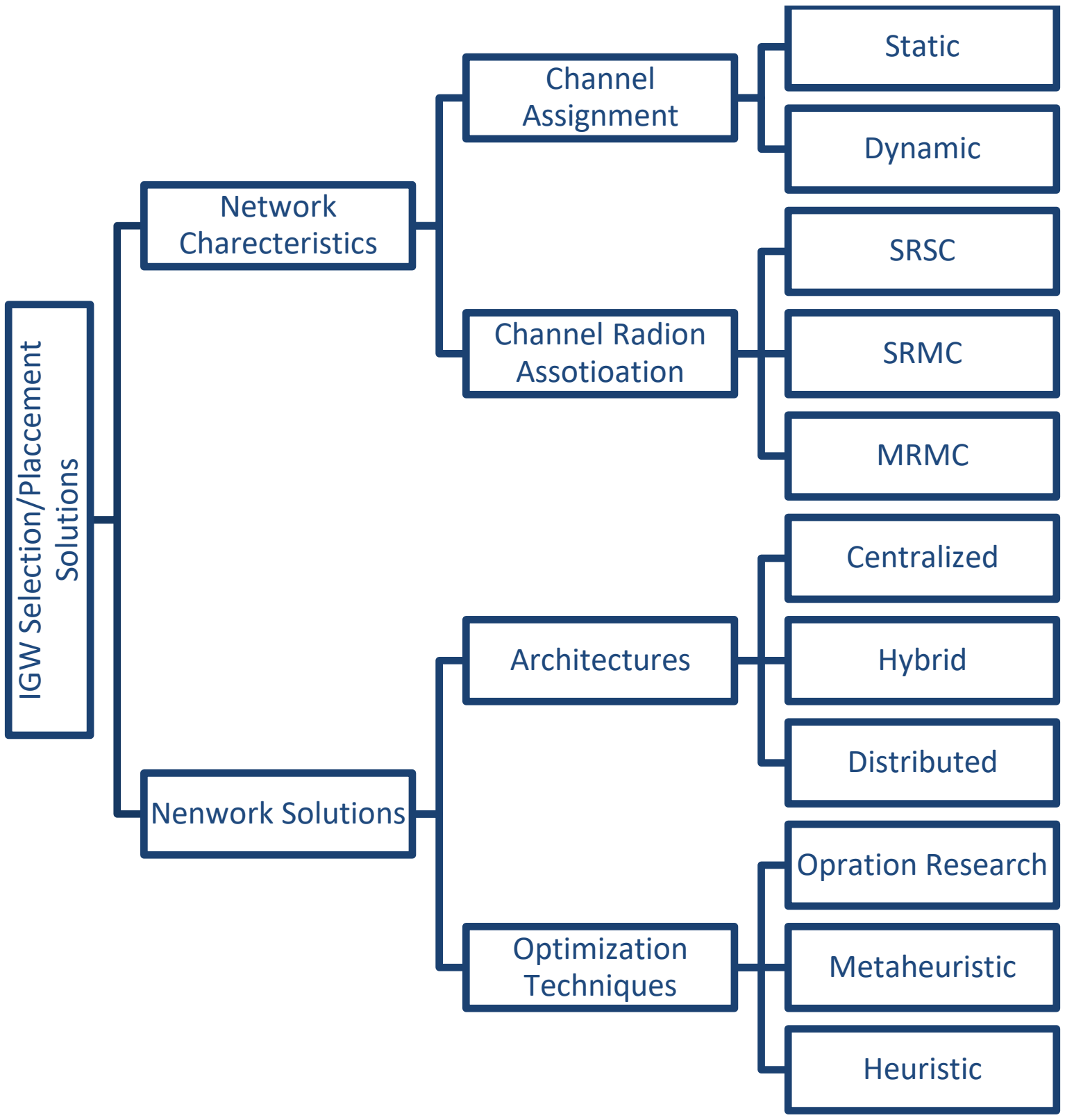

**Figure 2:** Different classifications in GPS protocols

### 3.A. Network characteristics

In this section, different works are categorized into two types of classifications related to network characteristics. These two types are network channel-radio association and static or dynamic channel assignment.

### 3.A.1. Classification with regard to channel-radio association

Mesh networks are usually designed to work on a single channel and a single radio. In a multi-radio mesh network, every node can be equipped with only a few radios. Radios operating in the same frequency band will interfere with radios close to them. Because there are only two frequency bands of 2.4 GHZ and 5.2 GHZ for use by 802.11, a node is limited for using only two radios[7]. It is also possible to equip nodes with multiple radios to work in multiple channels. In that case the allocation of channels to radios will be a problematic issue which will be discussed in the next section.

3.A.1.1. Single-Radio Single-Channel WMNs

In such a network each MR is equipped with only one radio tuned on one channel. SRSC networks operate in half-duplex mode and cannot transmit and receive simultaneously. As a result, increment of communication hops leads to decrement of end-to-end throughput on a single route. It means transmitting and receiving cannot occur concurrently and when a node is on receiving mode it can switch to transmitting mode only when a frame is fully received [7].

In [23] each MR was equipped with one radio and a single common channel and communication between nodes and users was done via a separate interface and channel.

Each MR in [34] was equipped with two virtual radio interfaces over one physical radio interface in which one was for backbone transmissions and the other one was for local communications. Therefore, local communications and backbone communications did not influence each other. Furthermore, MRs or MCs could receive packets from only one sender at a time. Many other studies like [5, 18, 21, 22, 25, 26, 29, 30, 35-44] employed SRSC networks as well.

3.A.1.2. Single-Radio Multi-Channel WMNs

MRs in this category are equipped with only one radio like the previous category but it is possible to use multiple channels. Therefore, interfering wireless links operate on different channels, enabling multiple parallel transmissions [45]. Because wireless interference is a critical restriction on applications of WMNs, applying channels at nodes next to each other for sending and receiving, can distinctly improve the throughput by decreasing interference [7].

A multi-channel WMN is assumed in [45] where interfering wireless links operate on different channels. Therefore, it enabled multiple parallel transmissions.

Among the works which were investigated only [45] has studied the GPS in WMNs using single radio. Most of the works tend to choose multi radios when they want to work on multi channels WMNs. Now, multiple channel networks with multi-radio MRs is investigated.

3.A.1.3. Multi-Radio Multi-Channel WMNs

One of the characteristics of WMNs is that communication channels are shared by the wireless terminals. Therefore, one of the major problems facing WMNs is the reduction of capacity due to interference caused by simultaneous transmission [26, 46]. Compared to omni-directional antennas, directional antennas provide spatial separation that can reduce interference drastically and consequently lead to increased network throughput [21, 47]. Moreover, due to the directional transmission of directional antennas they provide high antenna gain and as a result increase transmission range. Nevertheless, due to the cost and system complexity the number of directional antennas which can be installed at each node is limited. Figure 3 illustrates a MR MR which consists of “m” radios that each radio can be switched on “n” channels. These channels may be completely orthogonal or interfere with each other.

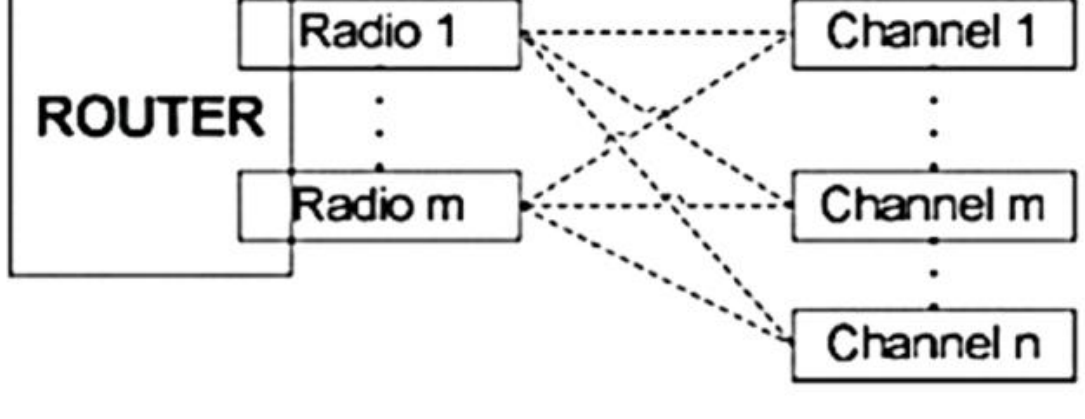

**Figure 3:** Multiple Radios Multiple Channels WMN [7]

As mentioned before, [26] considered the network with a single channel but the authors refined the linear programming for MRMC networks. Like [26] a MRMC network was considered in [24].

In [19], the authors presented a novel zero-degree algorithm for clustering the backbone WMN. In their study they worked on MRMC network. Several other works [27, 28, 31-33, 47, 48] also assumed multi radios in a multi-channel network in their study.

After investigating the GPS in the literature about the number of radios and channels we came to this conclusion that works proposed in early and mid-2000s considered SRSC mesh networks and due to the advantages of MRMC mesh networks, which was introduced later, such as the increase in the performance most papers recently focused on MRMC networks. Therefore, when employing a mesh network, it is more efficient to consider a MRMC network in the scenario.

### 3.A.2. Classification with regard to static and dynamic channel assignment

The performance of WMNs is influenced by many factors. Wireless multi-hop networks deal with an increasing number of users, which means the increment of demands. Interference between multiple contemporary data transmission (namely inter-flow and intra-flow interference) has a major role. To tackle this problem, the need for higher capacity is an issue. The reduction in throughput happens when two links communicate on the same frequency. Therefore, the number of users is decreased due to the interference.

Another issue is that MRs are working with one radio. Therefore, they cannot transmit and receive simultaneously. Equipping a wireless MR with multiple radios, in which each of the radios can work on different channels, is a practical solution. To provide several channels on a radio, it requires allocating particular channels to interfaces to reduce interference and maximize the network capacity. Channel dedication can be placed into two categories below:

#### 3.A.2.1. Static Channel Assignment

In static channel allocation, each interface of every MR is assigned a channel permanently. Therefore, interfaces do not switch channels and consequently they have lower overhead. However, they depend on stable and predictable traffic patterns in the network [46].

MRs in [27] were tuned to different channels for simultaneous transmission or reception. Like [27], MRs in [28] were equipped with two wireless interfaces to operate on separate orthogonal channel. The WMN backhaul architecture was divided in two layers. The regular WMN backhaul was used in the first layer for forwarding best effort traffic and an adaptive overlay scheme was presented in the second layer which connects the IGWs via intermediate relay MRs employing radio 2 for carrying delay sensitive traffic. Layer 1 and layer 2 operate independently since they are orthogonal for the contention free flow of data.

Similar to previous studies, in several other works such as [31-33, 48, 49], each MR was equipped with multiple radios tuned to different channels.

#### 3.A.2.2. Dynamic Channel Assignment

In this approach, an interface is allowed to switch from one channel to another channel frequently [46]. Thus, they have higher overhead than static strategies. When the network traffic changes frequently and when it is unpredictable, dynamic strategies are more appropriate. In the following, some papers in this category are investigated.

Authors in [24] considered the switching overhead in dynamic channel assignment during the transmission schedules in the scenario of IGW placement in MRMC wireless networks. In their approach they gave each link on a specific channel an interference-aware transmission schedule which assigned the time slot for transmission to maximize the overall network throughput meanwhile considering the switching overhead between the channels during the schedule.

Some works have considered both static and dynamic channel assignment like [26]. A greedy scheme was proposed in this work for SRSC WMN that could be extended to MRMC network as well. The algorithm satisfies the radio and channel constraints in both dynamic and static channel assignment.

In these works, balancing between network connectivity and increasing bandwidth and throughput were the main reasons the authors selected dynamic channel assignment in their works. Since the network traffic changes frequently by using this approach they are able to switch links between available channels.

### 3.B. Network Solutions

In this section, different works in GPS are classified into two main categories of centralized, distributed, or hybrid architectures and optimization techniques.

#### 3.B.1. Classification with regard to centralized, distributed or hybrid architectures

In this section, we aim at classifying the works in centralized, distributed, or hybrid approaches as follows.

3.B.1.1. Centralized architecture

One of the resources in every network is the IGWs. Placing the IGWs in wrong areas and selection of inappropriate nodes as IGWs may lead to underutilization. That is why resource allocation is of great importance in network. In centralized approaches, a central controller is responsible for all the nodes in the mesh network. This central node has all the necessary information about the nodes in the network. Therefore, this node is responsible for resource allocation and information distribution to other nodes according to its information. While centralized algorithms provide collision-free packet transmissions, but the number of applied routes is unnecessarily reduced, because a centralized algorithm uses a tree-based topology instead of a mesh-based topology[7]. An IGW-rooted tree is a directed graph as a tree and all edges are directed towards the IGW node, which is the root of the tree. Such network architecture offers benefits such as low routing overheads and efficient aggregation of flows and is suitable for IGW-oriented traffic because of simplifying the routing path from MRs to the IGW and increasing the network channel utilization.

Placing $k$ IGWs in the mesh network was the aim of [26] such that the total throughput that could be supported was maximized while considering fairness. The authors used a grid-based IGW deployment method. In this method at first they divided the whole deployment area into a grid and only placed the IGWs in the cross points on the grid. But the deficiency of the proposed IGW selection method is trying all the combinations of positions using linear programing and selecting the combination with the highest throughput. It also increases the computation cost since they used a lot of IGWs in their work. Therefore, the administrator needs to find an appropriate grid to satisfy both performance and cost requirements.

The main objectives on the IGW placement in [27] were minimization of number of IGWs and MR-IGW hop, and affordable computational complexity. In their approach, all MRs were associated with one IGW by using a tree structure. It was later proved that the search space of the linear program increases exponentially with the number of nodes. To solve this problem, the authors developed a heuristic algorithm for large-scale networks by dividing the graph into disjoint IGW-directed and connected clusters.

A centralized scheme was presented in [43]in which a coordinator was assigned to collect the current status of the neighboring IGWs. Thus, when a MC experiences a drop in the throughput, the coordinator sends a request including information about the throughput requirements and the current achievable throughput to neighboring IGWs having connectivity to the MC. This is a good idea, although collecting and sending the information to the IGWs takes plenty of time and therefore it increases the delay in the network.

The IGW-Load-Balancing algorithm that was proposed in [23] executes periodically in one of the IGWs, called the controller which can be any one of the IGWs. In case of the failure of the IGW, a new controller is elected from the rest of the IGWs. However, the election of IGW is not done fairly since it only selects the IGWs with lowest ID.

Lots of other articles such as [21, 24, 25, 29, 30, 35, 37-42, 49-51]concentrated on the centralized manner as well.

3.B.1.2. Distributed Architecture

In distributed algorithms, there is no node with global knowledge about the entire network. Instead, each node has a local knowledge of itself and its neighbors. In this approach, decisions are based on this local information and defined metrics. Thus, nodes communicate with each other by passing messages and collecting information[7].

In the routing mechanism which was proposed in [22], each MR constructs a routing table and each IGW broadcasts a message periodically. MRs receive the messages and update their table and select a primary IGW considering the traffic demand. This distributed idea for selecting an IGW is efficient however, a huge number of beacons must be sent among the nodes in order to update the tables for selecting the appropriate node as IGW.

In the proposed distributed algorithm in [36], the plane is partitioned into rectangles and for each rectangle the node with the largest ID is elected as the leader. After determining all nodes lying in the rectangle, the leader finds a solution for each rectangle. Union of the solutions found in rectangles is the solution. This is a distributed perspective which has a lower approximation ratio comparing to the other works.

3.B.1.3. Hybrid architecture

In this approach, the network is divided into clusters and each IGW as the head of the cluster each of which has the capability to satisfy the traffic demand of all MRs belonging to the cluster. Each cluster is connected without cyclic path. Each edge is directed from an MR toward the IGW and every MR is connected to the IGW by a single or multiple hop.

The authors in [45] aimed at placing a minimum number of IGWs. A Weighted Clustering Algorithm (WCA) was adapted for selection of cluster head. Those nodes whose neighbor number is less than a fixed threshold can be selected as a cluster head in WCA. The drawback is equal treating of cluster heads and normal nodes. Therefore, they presented a genetic algorithm to divide the WMN into clusters of bounded radius under relay load and cluster size constraints. In each cluster a spanning tree rooted at the IGW is used for traffic aggregation and forwarding. In [33], the clustering technique ensured a proper placement of IGWs leading to less deployment cost while providing enough network throughput capacity.

A zero degree algorithm was proposed in [19] for clustering based on number of MRs' connection. The graph is divided into disjoint clusters and each cluster has a head cluster aiming at minimizing the number of clusters and reducing the number of IGW-MR hops and distributing the IGW in the locations that are closest to available Internet connection points. The algorithm will be discussed in 3.B.2.2.

Similarly, in [18, 28, 31, 32, 34, 47, 48, 52], WMN was logically divided into a set of disjoint clusters and in each cluster a cluster head served as a IGW, connected directly to the network. A spanning tree rooted at the IGW was used in each cluster for traffic aggregation and forwarding. Each node was associated to one tree and would attach to another tree in case of path failure.

Both centralized and distributed algorithms were argued in [36] in which the aim was selecting a subset of nodes as IGW to serve the mesh nodes with the overall placing cost. They first introduced a centralized algorithm employing double partition which means partitioning the plane into large blocks, then partitioning each block into some squares. Then they extended it in a distributed manner.

After reviewing these works we have come to this point that most of the centralized solutions which were proposed are appropriate for small scale networks. Since if a central node has all the necessary information about the network as the network expands, collecting the information from all the nodes will be much more time consuming and although the decisions are more precise, the delay will also increase. On the other hand, for increasing the decision-making process and also the performance in larger scale networks, distributed perspectives are preferred. However, decisions are made based on local information and they might not be as precise and efficient as centralized approaches. To benefit the advantages of both approaches, most works have considered a hybrid perspective in which the network is centralized and distributed. Hybrid approaches are especially popular and great when a clustering scheme is proposed.

### 3.B.2. Classification with regard to Optimization Techniques

In this classification, the studies can be looked through in three groups of algorithms. Optimization based on operation research is the first group, the second class is dedicated to heuristic optimization schemes and the last one is meta-heuristic based optimization strategies. Different studies have used one of these methods according to their objectives.

#### 3.B.2.1. Operation research-based methods

In operation research-based methods, criteria such as interference, the number of available radios, the set of usable channels and other resource constraints at nodes are taken into account and are formulated [7]. In this subsection, the studies which considered this technique in their work are investigated. These works narrow down the solution space of the solution by defining some constraints in their single or multi-objective approach. In the following we will briefly introduce the objective and the constraints which are considered in the works.

The ideas proposed in [31], [32] and [52] were similar. The network was logically divided into a minimum number of disjoint clusters that cover all nodes and satisfy all the three QoS constraints which are cluster size, relay traffic, and cluster radius. Then, since the proposed formulation is not suitable for large-scale networks, they have proposed a heuristic which will be explained in subsection 3.B.2.2.

In [27], they modeled the IGW placement problem taking into account some constraints like full coverage, IGW throughput capacity, MR throughput capacity, interference, and investment cost. The boundaries which are set in the formulation are tree size, MR-IGW hops, and relaying load.

Having created clusters in [18], they used the routing tree rooted at the IGWs to discuss the QoS requirements in the backbone satisfying the sum of the local traffic on the nodes, the distance between any node, the relay traffic, the degree of each node, and the total number of nodes in the tree. Then they formulated the IGW placement problem to linear programming. The authors in [24] develop an approach in order to maximize the overall network throughput. To achieve this aim, they formulate the IGW selection problem in MRMC WMNs taking physical interference model and switching overhead into account.

The IGW placement problem as a linear program problem with multiple optimization objectives and some constraints were presented in [25]. The proposed approach achieved a good load balancing. However, no improvement was achieved in the number of IGWs and the hop count in their study. The main criteria which were considered in [33] and [33, 49] were network deployment cost, network throughput and congestion of IGWs. Some constraints were added to guarantee full coverage for MCs.

In [22], the IGW placement problem which involves reducing link interference and assuring fault tolerance was formulated as a problem with multi-objective ILP. The objectives of the formulation are minimizing the number of IGWs, minimizing the average MR-IGW hop count, and minimizing the IGW.

To evaluate the performance of IGW placement in [47], the problems of throughput optimization was formulized which led to two throughput metrics. The aggregate throughput and the worst-case per-client throughput were the metrics which were maximized. Some other works like [19, 26, 29, 30, 43] also proposed methods based on operation research in the same way.

There is a tradeoff between adding valid inequalities and the number of constraints. As the number of valid inequalities increases, the solution space will narrow down however as the number of constraints increases, it may cause clumsiness in terms of the running time of the model [4]. One of the essential parts of the proposed works is clustering. To formulate a solution for clustering nearly all of the works have considered an upper bound on the cluster size, relay traffic, and cluster radius. These three constraints are the most important factors needed for clustering.

After reviewing these works, we have come to this conclusion that operation research solutions are more precise. Nevertheless, due to high computational complexity and consuming more time comparing with heuristic and meta-heuristic solutions, they are more appropriate for small sized networks. Most of the works which proposed a solution for IGW selection or IGW placement provided a heuristic or meta-heuristic solution after the formulation in order to extend the scale of the network. Therefore, in this section most works have had virtually the same perspective and have taken into consideration the same constraints.

3.B.2.2. Heuristic-based Optimization Methods

When classic methods are too slow or fail to find any exact solution, heuristics are designed for solving the problems more quickly. Therefore, a heuristic process may include running tests and getting results by trial and error. As more sample data is tested, it becomes easier to create an efficient algorithm to process similar types of data. The algorithm may not be perfect or be the best of all the actual solutions to that problem but it is still valuable since it is not time-consuming. The objective of a heuristic is to give a solution that generates precise results in an acceptable amount of time.

Greedy algorithms are an example of this category. These algorithms look for simple and multi-step problems by deciding which next step will provide the most obvious benefit and make local optimal choice at each stage hoping the chosen step lead to a global optimum.

In the following we have investigated different perspectives in GPS which focused on heuristic-based methods. We have tried to briefly explain the idea and go over their weaknesses and strengths.

A greedy approach for approximating a Dominating Set (DS) is presented in [32]. At first the adjacency matrix is computed representing connectivity graph consisting of the DS of the previous iteration. Then a node that covers the greatest number of remaining nodes that are uncovered is selected iteratively. In this approach a chance is given to different feasible clusters to form before moving to the next iteration and increases the coverage of clusters. However, in this study the decision making is done greedily and a node that covers a maximum of uncovered nodes is elected.

The main idea of the scheduling in [26] was sorting the links depending on the interference models and then assigning the earliest time slot to a link that will not cause any interference to already scheduled links. Since they used a lot of IGWs in their work they achieved better throughput, connectivity, and coverage. On the other hand, as we mentioned before, the cost of the equipment was also increased.

In [27], a heuristic algorithm was developed including a degree-based GDTSP, which emphasized the connectivity degree of IGW meaning how many MRs were connected to the IGW, and a weight-based GDTSP, which placed emphasis not only on coverage but also on MR-IGW hop and selects more MRs close to the IGW. In degree-based algorithm, all nodes within R-hop are treated similarly in terms of connectivity while in weight-base method higher value is given to MRs with fewer hops. However, they have defined a formula for calculating the available bandwidth for each IGW when connecting an MR to it in the cluster phase. Whereas updating the table for the available bandwidth takes a lot of time.

Three heuristics were proposed in [42] to position a single IGW in WMN. Based on the minimum hop metric, they proposed their first heuristic. The second heuristic selects the IGW position considering the transmitter power. The third heuristic uses a shortest path algorithm to compute the minimum weight path from any node $n$ to the IGW positioned in $m$. Their second heuristic was sub-optimal and the last one selected a single IGW position and formed a better heuristic but this solution can only be used in a single IGW networks.

In the proposed heuristic in [52], each node which is searched by the IGWs joins the cluster whose IGW has the least load. If the input does not provide enough IGW nodes, Greedy Algorithm for Load Balancing Clustering (GA-LBC) will randomly select some of uncovered nodes as IGW nodes to build more clusters in order to cover all uncovered nodes. Using this method, the number of generated IGWs is nearly equal to other IGW placement approaches.

A Cost-Sensitive and Load-Balancing IGW Placement Algorithm (CSLBA) was presented in [18]. In IGW selection step, the nodes with high capacity to cost ratio have the high probability to be selected as IGWs. However, in this work only capacity and cost have been considered for selecting a IGW which seems not to be sufficient for a network with a lot of users.

A new approximation algorithm for IGW selection was proposed in [24] using a cross-layer throughput optimization exploiting the available resources. The new IGW placement scheme combining with their interference-free link channel scheduling had only a constant factor. Simulation result demonstrated that the proposed mechanism achieves much higher network throughput than random, fixed deployment and grid-based methods.

In [25], the authors proposed a two-stage algorithm that finds an optimal solution of IGW placement in a WMN. The first stage is a weight-based greedy IGW selection that selects the node with the maximal weight as a IGW and the second stage is a load balanced MR attachment in which the IGW with the minimum load should be given priority to make the node attached to. Simulation result shows that the algorithm has almost the same performance on both number of IGWs and the average MR-GW hop count compared to two existing approaches, while achieving better load balance.

A Backbone IGW Selection (BGS) scheme in [41] is presented. BGS contains an IGW and route selection and a proactive approach for IGW discovery. The IGW and route selection scheme was the combination of three metrics which are IGW load, interference, and expected link quality. However, in a large-scale network having multiple IGWs, the node requires more memory space to store routing information and spends more calculation for finding the best route to the IGW.

In [48], the author adapted an incremental clustering. The algorithm identifies IGWs based on the R-step transitive closure and assigns MRs to the identified IGWs iteratively. The deficiency of the algorithm is that using last step of transitive closure to select the IGWs might produce some nodes with zero connection. As a result, the nodes with zero connection should be selected as IGW in next iterations and this will lead to an increase in number of IGWs.

The proposed algorithm in [39] starts with identifying the relevant characteristics of paths to different IGWs by the MCs and then IGW selection is done. Considering a fairness constraint and non-IGW bottlenecks, the algorithm was designed to maximize the aggregate throughput of the network. For this solution it can be argued that by arbitrarily breaking the locality of the traffic in the network, contention will greatly increase and can perform worse than single nearest IGW association.

The unfairness issue was argued in [40]. They classified IGWs into high data rate IGWs, which are responsible for handling the traffic of the high data rate MCs, and low data rate IGWs, which are for the traffic of the low data rate MCs. However, it is not mentioned how they can find which areas need to be placed with high data rate IGWs and which with low data rate IGWs. Moreover, with this proposal it is of great importance to place the high data rate IGWs one hop away from the high data rate MCs since if they are two or three hops away the unfairness problem occurs again.

Due to the drawbacks of obtaining an optimal solution with a mixed-integer nonlinear programming in [23], the authors proposed an online approach which continually monitors network conditions and based on them switch sinks from congested to uncongested domains.

While, as mentioned before the election of IGW is not done and the proposed idea is centralized and as a result not appropriate for large scale networks.

An interference-aware IGW placement algorithm was presented in [22], in which the authors iteratively selected IGWs, and completed IGW association by generating IGW-rooted relay trees. However, in their work they only considered interference among IGWs, while MR-IGW and MR-MR interference was not taken into consideration. In [36], to select a subset of nodes as IGWs considering the cost two algorithms, which were centralized and distributed, was proposed which were discussed in 3.B.1.2 and 3.B.1.3.

An IGW placement scheme was proposed in [47] in which a multi-hop weight was calculated iteratively on the MRs, and each time a new IGW was placed on MR with highest weight. The weight computation takes the traffic flow into consideration with number of MRs, MCs, IGWs, traffic demand from MC, interference, and locations of IGW. But the deficiency of the algorithm is that their method is not updated in each step and IGW locations are discovered sequentially. As a result, the location of IGWs which were placed previously influences the location of IGWs placed later.

Given the traffic demands in [21], the authors proposed an IGW placement scheme that considers the capacity as a factor for selecting an IGW. However, in this work random traffic is designated to each node, some lower degree nodes may be assigned with heavy local traffic. Consequently, if they are selected as IGWs the average minimum and maximum hop count will increase.

In [28], a clustering algorithm for utilizing the stable links in a multi-hop WMN called OLSS was presented. Then they applied SMS algorithm in three steps of split, merge, and shift phase. Later, they proposed an Adaptive Overlay in which nodes will choose optimal one of the different possible paths. The adaptive overlay will carry types of traffic with minimum delay and with less contention and interference.

To satisfy the bandwidth demand of all the nodes a three step approach was presented in [35]. In the first step, the authors presented a graph partitioning and then they suggested an algorithm to ensure that the under loaded partitions share the load of their neighboring overloaded partitions. Finally, they defined the set of constraints to be observed while transiting a node to the wired network. In this work, the authors have reduced the time complexity from NP hard to a polynomial time complexity. The average delay in their work is lower than the two other methods and the average packet delivery ratio comparing to two other approaches is better.

The problem of throughput performance was highlighted in [51] and an IGW selection method was proposed. In this study, node with less total traffic in its collision domain is selected as IGW. But this algorithm might be problematic in case of high traffic in the network. Selecting IGWs in areas with little total traffic may lead the nodes with high traffic demand to connect to the nearest IGW with multiple hops and that may increase the delay.

In [19], two sub-algorithms called zero-degree(S) and zero-degree (L) formed their heuristic. The algorithms iteratively identify IGWs and assigns MRs to the identified IGWs taking node degrees into account. However, the IGW nodes in zero degree (S) are selected close to each other and in Zero degree (L) some IGWs are underused.

#### 3.B.2.3. Meta-Heuristic -based optimization methods

The mathematical solutions often lead to a formulation for a specific problem and are not scalable enough while there are other solutions that can be used for most of the problems. A meta-heuristic is a higher-level procedure or heuristic designed to find or select a lower-level procedure or heuristic that may provide a sufficiently good solution to an optimization problem, especially with incomplete or imperfect information or limited computation capacity. Compared to optimization algorithms and iterative methods, meta-heuristics do not guarantee that a globally optimal solution can be found on some class of problems. By searching over a large set of feasible solutions, meta-heuristics can often find good solutions with less computational effort than algorithms, iterative methods, or simple heuristics [53].

• **Simulated Annealing (SA):** SA algorithm is inspired by the cooling process of metals by which a material is heated and then cooled in a controlled way to increase the size of its crystals and reduce their defects. The heat causes the atoms to leave their initial positions and move randomly and the slow cooling gives them more likelihood to find configurations with lower energy than the previous one [37]. Annealing obtains the initial solution by randomly selecting IGW, and then the path of each of the node to the IGW is determined and assessed by the cost function to determine the best solution. Simulated annealing makes use of neighborhood exploration to obtain the optimal solution to the optimization problem[26]. In [45], a genetic algorithm and SA based algorithm was proposed which will be explained in next part.

An SA approach was presented in [37] to reach two maximization objectives, namely, network connectivity and user coverage. The algorithm starts by generating an initial solution. Then the fitness function follows a hierarchical approach. In the end, the solutions are compared and then accepted under some defined situations.

• **Genetic Algorithm (GA)**: GAs are numerical optimization algorithms inspired by both natural selection and natural genetics. They represent an intelligent exploitation of a random search used to solve optimization problems. GAs exploit historical information to direct the search into the region of better performance within the search space. The basic techniques of the GAs are designed to simulate processes in natural systems necessary for evolution. A GA algorithm attempts to find the best solution from a set of candidate solutions. A chromosome or solution is composed of several genes or variables and is generated from a genetic mutation and corresponds to a potential solution. In contrast to heuristic and greedy algorithms, the genetic algorithms do well in multiple goal optimization searching [54].

To optimize the objective in [45], which was selection of cluster head, the authors used GA and SA. In the first step, by generating randomly $L$ integer arrangements and computing their weight, $L$ sets of cluster heads with smaller weight are selected. Then, according to “accept or reject” criteria in SA if the weight of new $L$ sets is lighter, the old $L$ sets would be accepted. The main idea of WCA is determining the number and also location of IGWs. However, there is an assumption in this algorithm that locations of nodes are known, so that nodes can be selected as IGWs for cluster planning.

A Genetic IGW Placement Algorithm (GGPA) was proposed in [18] to achieve two goals of minimum cost and load balancing on the IGWs. The fitness function considers IGW placement cost and load variance on the IGWs in individual. Comparing the heuristic and the GA algorithm in their work, CSLBA leads to lower computing complexity and GGPA achieves better quality and had the advantage of global search for multiple goals. The genetic algorithm has the advantage of global search for multiple goals and has a better solution at the price of computing complexity.

A Hybrid Algorithm for Load Balancing Placement of IGWs (HA-LBPG) was proposed in [52]designing a GA based on the greedy algorithm GA-LBC. The fitness function was defined in a way that, the larger value of IGW number and deviation of load leads to a smaller value of fitness function. The GA-LBC was used for updating mutated individuals to denote the valid solutions of the IGW placement problem, and the updated individuals were put into the new generation population. Using this scheme, number of IGWs which are generated by HA-LBPG is nearly equal to results from other IGW placement approaches and its performance is better than other existing techniques.

To exemplify how different stages in GA work, we take a closer look at [34]. A IGW placement algorithm based on GA, PSO, and ACO was proposed in this work. At first, the MR with the highest weight was chosen as potential location for IGW placement using multi-traffic-flow weight. In the algorithm based on GA and PSO the fitness value of each scheme was calculated and step by step updated with the best method to quickly find the optimal. To have a better understanding of the crossover operator figure 4, which was also illustrated in their work, is shown. In cross over operator the algorithm begins with the pairs that include the individual with a higher fitness value until the population size becomes twice of the original size.

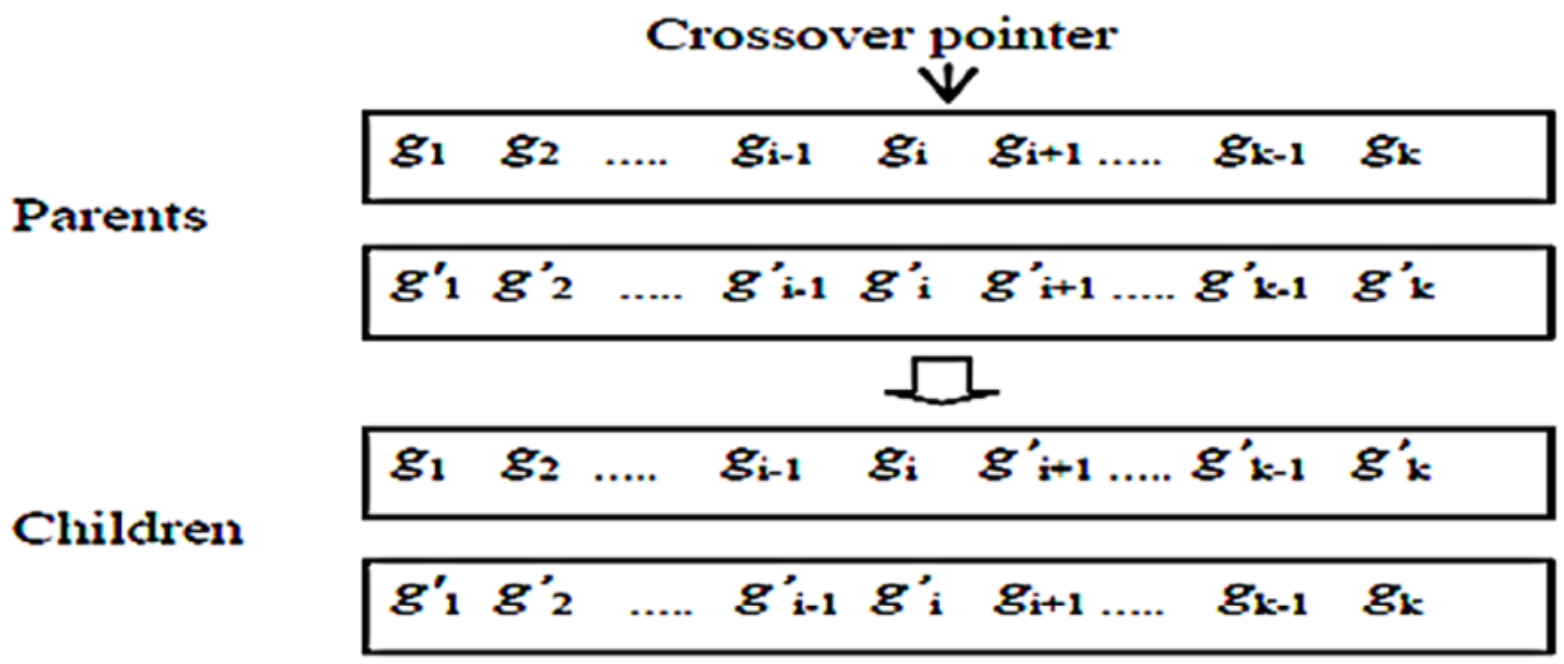

**Figure 4:** An example of crossover operator [34]

The process of random changing in the individuals in which one gene in the individual is changed with a certain probability is called mutation operator, as shown in Figure 5.

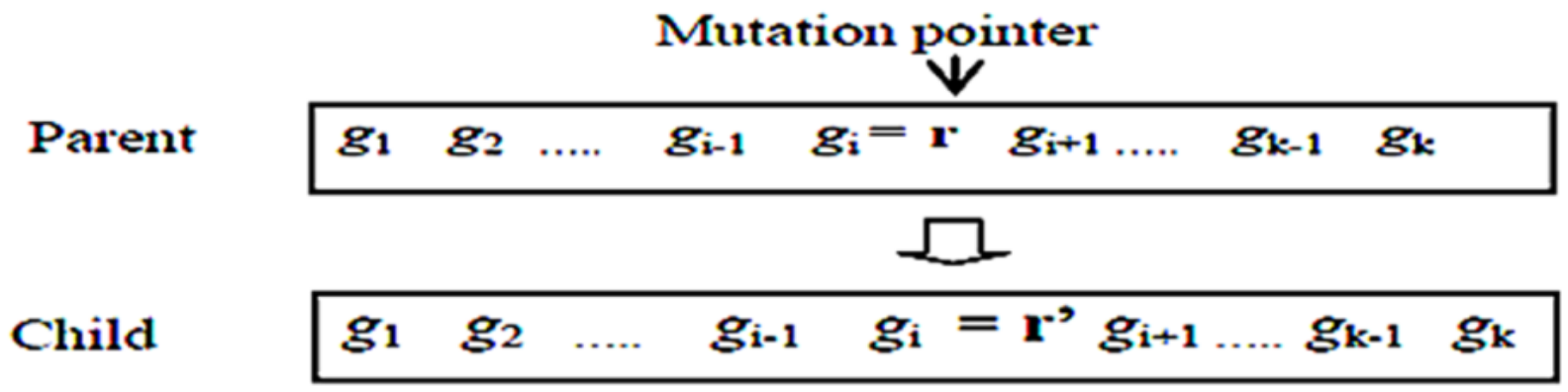

**Figure 5:** An example of mutation operator [34]

A GA method was proposed in [50] aiming at finding a location assignment for the MRs to the cells of the grid area that maximizes the network connectivity and MC coverage. They defined several specific mutation operators. When the mutation is done, network connections are re-computed. The simulation results proved the effectiveness of the proposed solution in terms of average size of giant components and average number of covered mesh MCs.

• **Particle Swarm Optimization (PSO):** PSO is a population-based stochastic optimization technique inspired by social behavior of bird flocking. PSO shares many similarities with evolutionary computation techniques such as Genetic Algorithms (GA). However, unlike GA, PSO has no evolution operators such as crossover and mutation. In PSO, a set of particles is placed in the search space of a given optimization problem and each particle fly through the problem space by following the current optimum particles. Then, each particle determines a move through the search space by combining the history of its own current and best locations with those of one or more particles of the swarm, with some random perturbations. Next iteration begins, after all particles have been moved [53].

As mentioned before a PSO method was also proposed in[34]. The initial population was generated with *P* element. Fitness value was calculated by a formula considering throughput of the MCs. In this algorithm for each particle, if the fitness value was better than the best fitness value in history this current value is selected.

To illustrate another PSO scheme in the literature we investigate the work proposed in [33]. The authors start by placing, for each particle in the swarm (a planning solution), a subset S1 of Access Points (Aps) to cover all Traffic Spots (TSs) (Coverage insurance design stage). Then there is a need to augment the set S1 by adding new MRs to connect the Aps together which consist of choosing the closest neighbor in one component graph to any node of a different component. Next, the shortest path between the two nodes is augmented and the algorithm stops when the final graph is connected. The same approach was used in [33, 49]. Figure 6 demonstrates the assignment and augmentation of Aps.

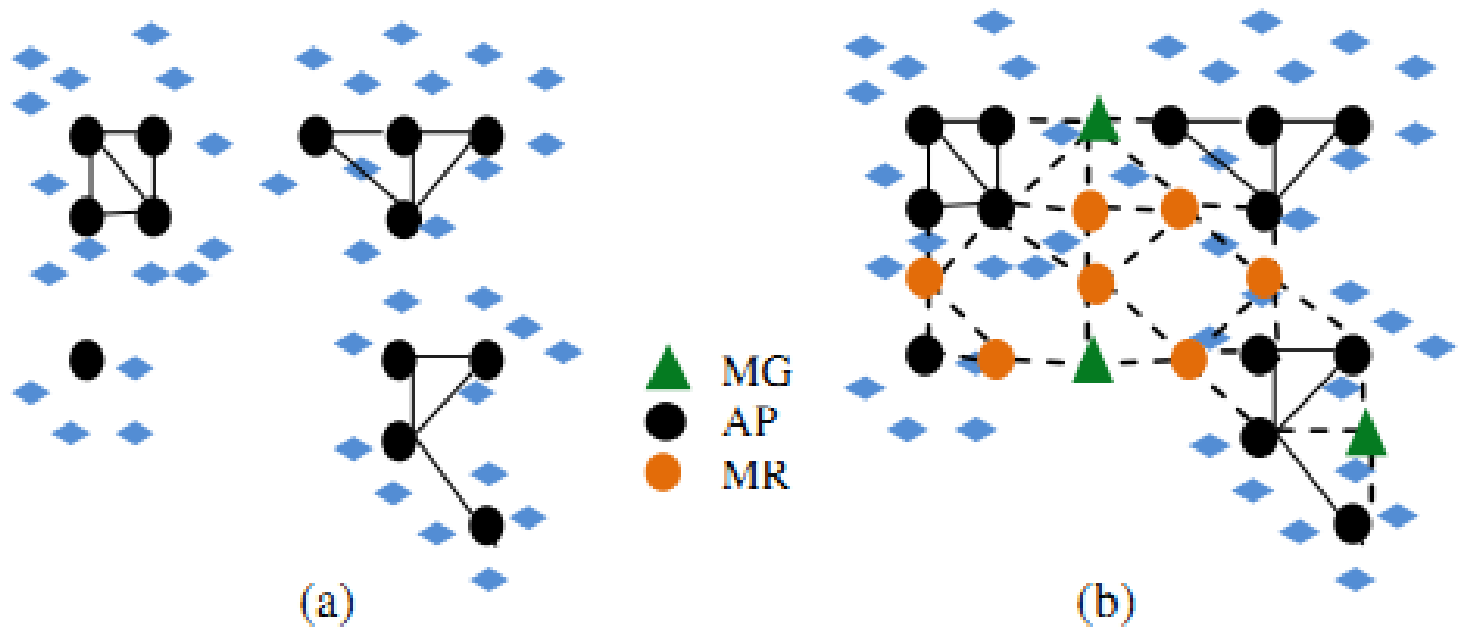

**Figure 6:** A feasible particle position: (a) TSs assigned (b) S1 augmented, MGs selected [33]

Experiment results show that, when CBGPA was coupled with the solution algorithm of the WMN design model, it did not only provide scalable and bounded delay planning solutions, but also in some instances, the solutions were cost-effective guaranteed.

• **Ant Colony Optimization(ACO):** As an excellent meta-heuristic based random algorithm, ant colony algorithm performs global optimization based on distribution by the positive feedback mechanism, continuously gathered and updated by pheromone so as to ultimately coverage to the optimal path [38]. Spontaneous creation and automatic layout are some of ant colony's characteristics.

A multi-path multi-IGW WMN any cast routing protocol based on an ant colony optimization was presented in [38] using routing protocol AOMDV. This protocol can adaptively find multiple paths with fewer hops and balanced load of WMN and achieve multi-path any cast routes effectively in large network load. It was shown that the proposed protocol can solve the congestion and get higher packet delivery ratio and lower average end to end delay.

In [34], the author calculated the probability and pheromone values of ants that will choose to go from current IGW *i* to next client *j*. In each iteration, the pheromone values were updated by all the number of ants that had reached the destination successfully and found the optimal solution and the probability move of the ants were calculated.

## 4. Discussion

WMNs is an emerging technology which has received research focus. It is a cheap, easy to implement, reliable, and efficient networking solution. In this study, some of the mostly related studies in the scope of GPS are classified in different perspectives such as centralized, distributed, or hybrid architecture, number of channels and radios, static and dynamic methods and optimization techniques.

To have a better understanding, an overall preview of various investigations in the literature is chronologically summarized in table 2. The parameters investigated in different works have been set in uniform metrics shown in table 1.

**Table 1:** Uniform metrics of the invested parameters

| | |
|---|---|
| *Traffic* | *Traffic flow, traffic load, relay traffic, flow conservation, flow balance* |
| *Network Capacity* | *IGW capacity, link capacity* |
| *Interference* | *Interference* |
| *Delay* | *Delay, hop count* |
| *Cost* | *Investment cost, IGW placement cost, Deployment cost, overall cost* |
| *Throughput* | *Throughput capacity* |
| *Fairness* | *Fairness* |
| *Load balancing* | *Load balancing, Link load, IGW load* |
| *Congestion* | *Congestion, Bottleneck* |
| *Packet loss* | *Packet loss, packet delivery, loss ratio, link quality* |
| *Fault tolerance* | *Fault tolerance* |
| *User Coverage* | *User coverage* |
| *Topology* | *Network size, node degree, connectivity degree, number of MRs and MCs* |

**Table 2:** An overall preview of some of the mostly related studies

| Reference | Heuristic Meta-heuristic Operation research | Centralized Distributed Hybrid | Network Structure | IGW placement/ Selection | Investigated Parameters |
|---|---|---|---|---|---|
| [32] | Operation research, Heuristic | Hybrid | MRMC | IGW Placement | traffic, delay, network capacity |
| [55] | Meta-heuristic (GA) | Hybrid | multi-interface | IGW Selection | network reliability, traffic |
| [56] | heuristic | centralized | SRSC | IGW Selection | network reliability, traffic, throughput |
| [44] | Heuristic | Centralized | SRSC | IGW Selection | topology, load balancing, delay |
| [57] | Operation research, Heuristic | Centralized | SRSC | IGW Selection | interference |
| [45] | Meta-heuristic (GA, SA) | Hybrid | SRMC | IGW Placement | traffic, network capacity, cost, topology, power |
| [26] | Operation research, Heuristic | Centralized | MRMC SRSC | IGW Placement | throughput, interference, fairness |
| [27] | Operation research, Heuristic | Hybrid | MRMC | IGW Placement | delay, topology, cost, throughput, interference |
| [31] | Operation research | Hybrid | MRMC | IGW Placement | throughput, network capacity, delay |
| [43] | Operation research | Centralized | SRSC | IGW Selection | throughput, congestion |
| [42] | Heuristic | Centralized | SRSC | IGW Placement | delay, throughput, power |
| [52] | Operation research, Heuristic, Meta-heuristic(GA) | Hybrid | SRSC | IGW Placement | delay, network capacity, interference, load balancing |
| [18] | Operation research, Heuristic, Meta-heuristic(GA) | Hybrid | SRSC | IGW Placement, Selection | cost, load balancing |
| [24] | Operation research, Heuristic | Centralized | MRMC | IGW Placement and Selection | throughput, interference, fairness, traffic |
| [25] | Operation research, Heuristic(greedy) | Centralized | SRSC | IGW Placement, Selection | load balancing, delay |
| [30] | Operation research | Centralized | SRSC | IGW Placement | interference, traffic, IGW capacity, network capacity |
| [33] | Operation research, Meta-heuristic(PSO) | Hybrid | MRMC | IGW Placement | cost, congestion, interference, traffic |
| [41] | Heuristic | Centralized | SRSC | IGW Selection | load balancing, interference, packet loss |
| [39] | Heuristic(greedy) | Centralized | SRSC | IGW Selection | fairness, throughput, congestion |
| [40] | Heuristic | Centralized | SRSC | IGW selection | fairness, throughput |
| [48] | Heuristic | Hybrid | MRMC | IGW Placement | delay, load balancing, network capacity |
| [33, 49] | Operation research, Meta-Heuristic(PSO) | Centralized | MRMC | IGW Placement | cost, congestion, interference, traffic |
| [23] | Operation research, Heuristic | Centralized | SRSC | IGW Selection | congestion, load balancing |
| [22] | Operation research, Heuristic | Distributed | SRSC | IGW Placement | interference, fault tolerance, delay |
| [38] | Meta-heuristic(ACO) | Centralized | SRSC | IGW Selection | load balancing, packet loss, delay |
| [37] | Meta-heuristic(SA) | Centralized | SRSC | IGW Placement | cost, user coverage, topology |
| [36] | Heuristic | Centralized, distributed | SRSC | IGW Placement | cost |
| [47] | Operation research, Heuristic | Hybrid | MRMC | IGW Placement | delay, interference, topology |
| [21] | Heuristic | Centralized | SRSC | IGW Placement | Traffic, delay |
| [29] | Operation research | Centralized | SRSC | IGW Placement | interference, traffic |
| [28] | Heuristic | Hybrid | MRMC | IGW Placement | delay, network capacity, load balancing |
| [35] | Heuristic | Centralized | SRSC | IGW Placement | load balancing |
| [51] | Heuristic | Centralized | SRSC | IGW Selection | throughput, congestion, traffic |
| [19] | Operation research, Heuristic | Hybrid | MRMC | IGW Placement | delay, load balancing, network capacity, delay |
| [34] | Meta-heuristic (GA,PSO,ACO) | Hybrid | SRSC | IGW Placement | load balancing, cost |
| [50] | Meta-heuristic (GA) | Centralized | SRSC | IGW Placement | topology |

## 5. Conclusion and future works

In this article, the existing works on GPS in WMNs were surveyed. We have investigated related works applying multiple radios and multiple channels in their works to increase overall throughput and capacity of network. Different works in SRSC, SRMC, and MRMC WMNs were surveyed. Channel dedication, which includes static channel assignment and dynamic channel assignment, was discussed. Later, channel assignment was classified into centralized, distributed, and hybrid architectures. Then a classification for GPS was highlighted, including heuristic, meta-heuristic based and operation research-based techniques.

Having investigated a wide range of works it can be noted that heuristic and meta-heuristic-based approaches provide an acceptable performance while the operation research based methods result in optimized solutions. In fact, in operation research-based solutions, QoS parameters are taken into account and formulated but they are not scalable enough. While soft computing solutions have the potential to improve and be used in extended networks, they obtain near-optimal results in polynomial times. Therefore, they are more appropriate for larger instances. Furthermore, centralized methods typically result in better performance in small networks and for large-sized network instances, distributed methods represent optimized solutions in reasonable times.

As previously discussed in this work by IGW selection we have surveyed papers which have considered to select a mesh node for being given the IGW functionality. There are many other works which have considered IGW selection as selecting next IGW for transmitting the packet, which is most related to routing. Due to the high number of existing articles about these scopes we were not able to study all of them in this survey. As a result, in the future we would like to study different approaches in this scope and the articles about routing, since they are interconnected, and classify the works in the literature considering some new categories.

Moreover, after investigating loads of works in the literature we have found that more parameters can be considered when proposing an idea in GPS. Multi-rate schemes lead to better throughput in the network. This important issue can be considered. Besides, due to minimal mobility of mesh MRs few works have considered energy consumption as one of the parameters in the simulation. Whereas, green MRs and the necessity of connecting to UPS for some MRs are convincing reasons to support the notion that energy consumption is also in demand in WMNs. In the future, we would like to propose an IGW *selection* and clustering scheme in a MRMC WMNs considering these parameters.